\documentclass[a4paper]{jpconf}
\usepackage{graphicx}

\usepackage{amsmath,amssymb,amsfonts,xspace}

\numberwithin{equation}{section}
\renewcommand{\Im}{\imag}

\newcommand{\be}{\begin{equation}}
\newcommand{\ee}{\end{equation}}
\newcommand{\beq}{\begin{eqnarray}}
\newcommand{\eeq}{\end{eqnarray}}
\newcommand{\ben}{\begin{eqnarray}\displaystyle}
\newcommand{\een}{\end{eqnarray}}
\newcommand{\bea}[2]{\be\label{#2}\begin{array}{#1}}
\newcommand{\eea}{\end{array}\ee}

\def\Tr{\,{\rm Tr}\, }
\def\det{\,{\rm det}\, }

\def\sign{{\rm sign}}

\def\Im{\,{\rm Im}\, }

\def\({\left(}
\def\){\right)}
\def\[{\left[}
\def\]{\right]}

\def\11{1\!\! 1}

\def\bi{\bar {\i}}

\newcommand{\de}{\mathrm{d}}

\newcommand{\I}{\mathrm{i}}
\newcommand{\cA}{\mathcal{A}}
\newcommand{\cB}{\mathcal{B}}
\newcommand{\cI}{\mathcal{I}}
\newcommand{\cL}{\mathcal{L}}

\newcommand{\cH}{\mathcal{H}}

\newcommand{\cC}{\mathcal{C}}
\newcommand{\cS}{\mathcal{S}}
\newcommand{\cG}{\mathcal{G}}
\newcommand{\cK}{\mathcal{K}}
\newcommand{\cM}{\mathcal{M}}
\newcommand{\cW}{\mathcal{W}}
\newcommand{\cN}{\mathcal{N}}

\newcommand{\cX}{\mathcal{X}}

\newcommand{\IR}{\mathbb{R}}
\newcommand{\IC}{\mathbb{C}}
\newcommand{\IZ}{\mathbb{Z}}

\newcommand{\kscom}[1]{\kappa(#1)}

\def\varpi{t}

\newcommand{\nn}{\nonumber}
\newcommand{\kahler}{{K\"ahler}\xspace}

\def\bse{\begin{subequations}}
\def\ese{\end{subequations}}

\def\qli2{{\bf E}}

\newcommand\bOm{\bar\Omega}

\def\ea#1\ea{\begin{align}#1\end{align}}

\usepackage{graphicx}
\usepackage[arrow,curve,matrix,arc]{xy}

\newcommand{\Om}{\Omega}

\begin{document}
\title{Four ways across the wall}

\author{Boris Pioline}

\address{ Laboratoire de Physique Th\'eorique et Hautes
Energies, \\ 
CNRS UMR 7589 and Universit\'e Pierre et Marie Curie - Paris 6,\\
4 place Jussieu, 75252 Paris cedex 05, France}

\ead{pioline@lpthe.jussieu.fr}

\begin{abstract}
An important question in the study of $\cN=2$ supersymmetric string or field 
theories is to compute the jump of the  BPS spectrum across walls of marginal 
stability in the space of parameters or vacua. I survey four apparently 
different answers for this problem, two of which are based on the
mathematics of generalized Donaldson-Thomas invariants (the 
Kontsevich-Soibelman and the Joyce-Song formulae), while the other two are based
on the physics of multi-centered black hole solutions (the Coulomb branch
and the Higgs branch formulae, discovered
in joint work with Jan Manschot and Ashoke Sen \cite{Manschot:2010qz}).  
Explicit computations indicate that these 
formulae are equivalent, though a combinatorial proof is currently 
lacking.  
\vspace*{2mm}
\\ 
{\it Contribution to the proceedings of the workshop ``Algebra, Geometry and
Mathematical Physics",  Tj\"arn\"o 
Marine Biological Laboratory, Sweden, 25-30 October 2010}
\end{abstract}

\section{Introduction}

In quantum field theories and string theory vacua with extended supersymmetry, 
it is often possible to determine the spectrum of  BPS bound states in some 
weakly coupled region of moduli (or parameter) space $\cB$. In extrapolating the BPS 
spectrum to strong coupling, one usually faces two issues: i) short BPS multiplets 
may pair up into long multiplets and leave the BPS spectrum and ii) single-particle 
bound states may decay into the continuum of multi-particle states. The 
first issue can be avoided by considering a suitable index $\Omega(\gamma,t)$, 
designed such that contributions from long multiplets cancel. $\Omega(\gamma,t)$ is then 
a piecewise constant function of the charge vector $\gamma$ and couplings/moduli $t\in \cB$.

\begin{figure}
\centerline{\includegraphics[height=4cm]{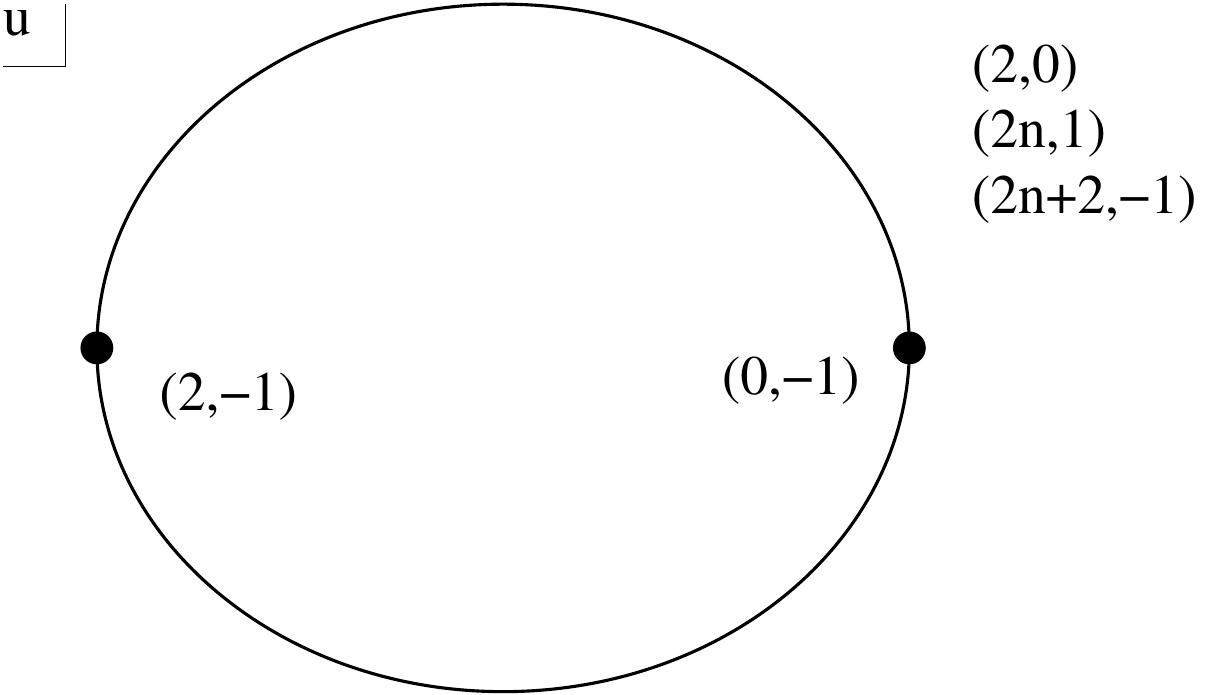} }
\caption{Chamber structure of the $u$-plane and BPS spectrum 
in $\cN=2, D=4$ SYM theory with $SU(2)$  gauge group and no flavor. 
The line $\Im(a/a_D)=0$
separates the strong and weak coupling chambers. The only
stable BPS states in the strong coupling chamber are the monopole
and dyons with charges  $(q,p)=\pm(0,1)$, $\pm(2,-1)$, in the conventions of 
\cite{Seiberg:1994aj}. The weak coupling
spectrum consists of these same states and their images around the monodromy
at infinity, plus the $W$-boson with charge $(2,0)$.  
\label{fig_sw0}}
\end{figure}

The second issue arises at certain loci in moduli space,
where the bound state becomes unstable towards decay into a $n$-particle 
state with charges $\{\alpha_i\}$ such that $\gamma=\sum_{i=1\dots n} \alpha_i$. In 
four-dimensional field or string theories with $\cN=2$ supersymmetry, 
the mass of a BPS bound state $M(\gamma,t)$ is equal to $|Z(\gamma,t)|$, 
where the central charge
$Z$ is a map from $\cB$ to ${\rm Hom}(\Gamma,\IC)$, where
$\Gamma$ is the charge lattice. In particular, $Z$ is linear in its first argument 
$\gamma$. The decay is 
therefore energetically possible only when 
(and even then, marginally so) the phase of $Z(\gamma,t)$ aligns with 
the phase of each of the $Z(\alpha_i,t)$'s, so that $M(\gamma)=\sum_{i=1\dots n} M(\alpha_i)$. 
This alignment takes place in a locus of codimension $p-1$ in the moduli space $\cB$, 
where $p$ is the dimension of the subspace
of $\Gamma$ spanned by the $\alpha_i's$. The dangerous case is $p=2$, where the 
locus defines a codimension one ``wall of marginal stability'' in $\cB$, across which the 
index $\Omega(\gamma,t)$ may jump.  A paradigm of this phenomenon
is Seiberg-Witten theory with $SU(2)$ gauge group and no flavors:  across the curve $\{ a/a_D\in \IR^+\}$ in the $u$-plane, the BPS spectrum jumps 
 from an infinite number of states in the weak coupling region
 to just two states in the strong coupling region -- the monopole and the dyon 
 (see Fig. \ref{fig_sw0})
 \cite{Seiberg:1994rs,Ferrari:1996sv}. An important physical question is therefore to determine 
 the jump 
 \be
 \Delta\Omega(M\gamma_1+N\gamma_2)=\Omega^-(M\gamma_1+N\gamma_2)-
 \Omega^+(M\gamma_1+N\gamma_2)
 \ee
 of the BPS index
  \footnote{Here, $\Tr'$ denotes
 the trace in the Hilbert space associated to $(\gamma,t)$ with the center of motion
 degrees of freedom removed, and $(-1)^F$ denotes the fermionic parity, equal
 to $(-1)^{2J_3}$ by the spin-statistics relation, where $J_3$ is 
 the angular momentum operator
 along the $z$ axis.}  $\Omega(\gamma,t)\equiv \Tr'(-1)^{F}$
 across the wall of marginal stability
 \be
 \label{defwall}
 W(\gamma_1,\gamma_2) = \{ t\in \cB \, / \, \arg[Z(\gamma_1,t)]=\arg[Z(\gamma_2,t)]\}\ ,
 \ee
 in terms of the BPS indices on side of the wall, say the 
 $\Omega^+(\gamma)'s$. Here we denoted by $\Omega^\pm(\gamma)$  the index in the chamber $c^\pm$ on the side of the wall 
 where $\arg Z(\gamma_1,t) \gtrless \arg Z(\gamma_2,t)$.
It will be convenient to  choose the basis $\gamma_1,\gamma_2$ such that 
 $\Omega^+(M\gamma_1+N\gamma_2)$ vanishes whenever $MN<0$ (the `root
 basis' condition \cite{Andriyash:2010qv}), and denote
 by $\tilde\Gamma=(\IZ^+\gamma_1+\IZ^+\gamma_2)\backslash \{0\}$ the positive cone
 in the two-dimensional sublattice spanned by $\gamma_1,\gamma_2$.  We further
 assume that $\langle \gamma_1,\gamma_2\rangle < 0$. 
 
 As we shall explain in \S\ref{sec_sugra}, in $\cN=2$ supergravity and for suitably 
 large charges, the jump in  
 $\Omega(M\gamma_1+N\gamma_2,t)$ is accounted 
 by the loss or gain of a family $n$-centered BPS black hole solutions with 
 charges $\alpha_i=M_i\gamma_1+N_i \gamma_2\in\tilde\Gamma$, 
 with $\sum_i(M_i,N_i)=(M,N)$, 
 which exist on the side $c_-$ of the wall, and on this side only. Close to the wall, 
 the $n$-centered configuration is loosely bound, with the relative distances $r_{ij}$
between the centers diverging at the wall. As a result, its index factorizes into
the product of the internal index $\Omega(\alpha_i)$ associated to each center,
times the index of the degrees of freedom associated to the relative motion of the
centers (with suitable modifications due to Bose-Fermi statistics when some of 
the $(M_i,N_i)$ coincide) \cite{Manschot:2010qz}. 
In \S\ref{sec_sugra}, we shall compute this configurational 
index using localization techniques, and obtain $ \Delta\Omega(M\gamma_1+N\gamma_2)$
for arbitrary values of $(M,N)$. 

A closely similar problem arises in the mathematics of Donaldson-Thomas invariants 
of coherent sheaves on a compact complex manifold $\cX$. These invariants, which 
we shall again denote by $\Omega(\gamma,t)$, are labelled by a class $\gamma$ in 
the K-theory lattice $\Gamma=K(\cX)$, and depend on a choice of stability condition 
$\phi_t:\Gamma\to S^1$ inside a complex family parametrized by $t$.  
$\Omega(\gamma,t)$ is defined as the Euler characteristic \footnote{Rather, the 
Euler characteristic weighted by Behrend's function, see 
\cite{MR2600874,Joyce:2008pc} for details.}
 of the moduli space $\cM(\gamma,t)$
of stable coherent sheaves on $\cX$ in the class $\gamma$ with 
respect to $\phi_t$. As the stability condition 
is varied, some of the stable sheaves
may become unstable and the DT invariant $\Omega(\gamma,t)$ may jump. 
This happens on the same walls of marginal stability as in \eqref{defwall}, where 
$\phi_t(\gamma)$ plays the role of $\arg Z(\gamma,t)$. The similarity between 
these two wall-crossing problems follows from the fact that stable 
objects in the derived category of coherent sheaves on a  Calabi-Yau three-fold
$\cX$ are realized physically by  BPS states in type IIA 
string theory compactified on $\cX$ (see e.g. \cite{Douglas:2002fj,Aspinwall:2004jr} 
for reviews).

In two independent pieces of work, Kontsevich-Soibelman \cite{ks}
and Joyce-Song \cite{Joyce:2008pc} have determined the 
variation $\Delta\Omega(\gamma)$ in terms of the DT invariants 
on one side of the wall. In both works, it was noted that the wall-crossing formula
takes a simpler form in terms of the `rational DT invariants' $\bar\Omega(\gamma,t)$,
related to the ordinary, integer-valued invariants $\Omega(\gamma,t)$ by 
the `multi-cover formula'
\be
\bar\Omega(\gamma,t)\equiv \sum_{m|\gamma} \Omega(\gamma/m,t)/m^2\ ,
\ee 
where the sum runs over all integers $m\geq 1$ such that $\gamma/m\in \Gamma$
(thus,  $\bar\Omega(\gamma)=\Omega(\gamma)$ if $\gamma$ is primitive). From
a physics point of view, we shall 
explain in \S\ref{sec_stat} that  the 
replacement $\Omega(\gamma,t)\to \bar\Omega(\gamma,t)$ effectively converts
the Bose-Fermi statistics of the $n$ centers into Boltzmannian statistics, 
thereby allowing us to treat the centers as distinguishable. 
In more detail, the KS and JS formulae 
express the jump $\Delta\bar\Omega(\gamma)$ as
\be
\label{jswcf2}
\Delta\bOm(\gamma,t) =
\sum_{n\geq 2 }\, 
  \sum_{\substack{ \{\alpha_1,\dots \alpha_n\} \in \tilde\Gamma\\
\gamma= \alpha_1+\dots +\alpha_n}}\, 
\frac{g(\{\alpha_i\})}{\prod m_k!}
 \prod\nolimits_{i=1}^n \bOm^+(\alpha_i,t) \ ,
 \ee
 where the second sum runs
over all (unordered) decompositions of the total charge vector $\gamma$
into a sum of $n$ vectors $\alpha_i\in\tilde\Gamma$. The coefficients 
$g(\{\alpha_i\})$ (after extracting out a Boltzmann-Gibbs symmetry factor $\prod m_k!$ 
whenever $\{\alpha_i\}$ contains $m_1$ copies of $\beta_1$, 
$m_2$ copies of $\beta_2$, etc) are universal functions of the $\alpha_i$'s. As we shall see,
the function  $g(\{\alpha_i\})$ turns out to be equal to the index $\Tr'(-1)^{F}$ of 
the configurational degrees of freedom of an $n$-centered Boltzmann 
black hole solution. To compute this index, it is convenient to consider
the  refined (or equivariant) index  $g(\{\alpha_i\},y)=\Tr'(-y)^{2J_3}$, 
evaluate the latter  by localization methods, and set $y=1$ at the end. In fact,
the refined configurational index $g(\{\alpha_i\},y)$ for general $y$ enters 
the wall-crossing formula for so-called motivic 
Donaldson-Thomas invariants, see \eqref{efirst3}. 

In the remainder of this survey, we first give an executive summary of the KS (\S\ref{sec_ks}) 
and JS (\S\ref{sec_js})  wall-crossing 
formulae, silencing the subtleties involved in defining the Donaldson-Thomas invariants
by themselves. In \S\ref{sec_wcbh} we then 
derive the combinatorial factors $g(\{\alpha_i\})$ by quantizing
the phase space of multi-centered BPS black holes and evaluating the index by 
localization. In \S\ref{sec_qui}, we give an alternative computation of $g(\{\alpha_i\})$
relying on Reineke's results for quivers without closed loops. 
We end in \S\ref{sec_conclu} with a discussion of some open problems. 
The material is 
mostly based on \cite{Manschot:2010qz,MPS2-in-progress}, which the reader should consult for more details. Other important references include \cite{Denef:2000nb,Denef:2002ru,Denef:2007vg,Gaiotto:2008cd,deBoer:2008zn,ks,Joyce:2008pc,Dimofte:2009bv,Dimofte:2009tm}.

\section{The Kontsevich-Soibelman wall-crossing formula\label{sec_ks}}

We start by reviewing the Kontsevich-Soibelman wall-crossing formula for 
generalized Donaldson-Thomas invariants. For generality, we first state its 
motivic (aka refined) version, and then discuss its classical limit. As an application, 
we use the KS formula to rederive the primitive and semi-primitive wall-crossing formula. Finally, 
we extract the coefficient $g(\{\alpha_i\})$ appearing in \eqref{jswcf2} for $n\leq 3$,  
for comparison with other wall-crossing formulae discussed later.

\subsection{The motivic KS formula}

The motivic KS formula pertains to `motivic Donaldson-Thomas invariants'
$\Omega_n(\gamma,t)$ attached to a Calabi-Yau threefold category $\cX$,
with a stability condition $\phi_t$. Informally, $\Omega_n(\gamma,t)$ is
the  $n+\frac12 d$-th Betti number of the moduli space $\cM(\gamma,t)$ of stable 
degree-$\gamma$ objects in the triangulated category of coherent sheaves 
on $\cX$, where $d=\dim \cM(\gamma,t)$. We define the Poincar\' e  polynomial 
\be
\label{refdeg}
\Omega_{\rm ref}(\gamma,t,y) = 
\sum_{n\in \IZ} (-y)^n \, 
\Omega_{n}(\gamma,t) \, ,
\ee
which is a finite Laurent polynomial in $y$, symmetric under $y\to 1/y$ (the subscript `ref'
stands for 'refined', which for our purposes 
is synonymous with `motivic' \cite{Dimofte:2009bv}).

To state the KS formula, we introduce the Lie algebra $\cA$ spanned by abstract 
generators $e_\gamma$, for each $\gamma\in K(\cX)=\Gamma$, subject to 
the commutation rule
\be
\label{KSalg}
[ e_{\gamma} , e_{\gamma'} ] =  
\kscom{\langle \gamma,\gamma'\rangle,y}\, e_{\gamma+\gamma'},
\ee
where $\langle \gamma,\gamma'\rangle$ is the integer-valued antisymmetric 
pairing on $\Gamma$ (physically, the Dirac-Schwinger-Zwanziger product
for electromagnetic charge vectors), and 
\be
\label{defkappa}
 \kappa(x,y) \equiv \displaystyle{\frac{ (-y)^{x} - (-y)^{-x}}{y-1/y}}
= (-1)^x \frac{\sinh(\nu\, x)}{\sinh\nu}, 
\qquad \nu\equiv \ln y \ . 
\ee
It is straightforward to check that \eqref{KSalg} satisfies the Jacobi identity for any $y$. 

Now, for a given choice of stability condition $t$ and any $\gamma\in \Gamma$ , let
 $U_\gamma(t)$ be the element in the  group $\cG=\exp(\cA)$ defined by
\be
\label{defU}
U_\gamma(t) = \prod_{n\in\IZ} \qli2\left( \frac{y^n\, e_\gamma}{y-1/y} 
\right)^{(-1)^{n+1}  
\Omega_{n}(\gamma,t)}\ ,
\ee
where $\qli2(x)$ is the quantum dilogarithm function
\be
\qli2(x) \equiv  \exp\left[ \sum_{k=1}^{\infty} \frac{(x y )^k}{k(1-y^{2k})}\right]\ .
\ee

We now restrict to $\gamma\in\tilde\Gamma$, where $\tilde\Gamma$ is the positive
cone in the two-dimensional sublattice of $\Gamma$ spanned by two primitive
vectors $\gamma_1, \gamma_2$, and assume that the 'root basis'
condition stated below \eqref{defwall} holds. 
The motivic KS wall-crossing formula \cite{ks,Dimofte:2009bv,Dimofte:2009tm} 
states that the following two ordered products
\be \label{ewallfinU}
\prod_{\substack{M\geq 0,N\geq 0, \\  M/N\downarrow}}
U^+_{M\gamma_1+N\gamma_2}
=
\prod_{\substack{M\geq 0,N\geq 0, \\  M/N\uparrow}}
U^-_{M\gamma_1+N\gamma_2}\, ,
\ee
where the products are ordered with decreasing (resp., increasing) values
of $M/N\in [0,+\infty]$ (such that the argument of $Z(\alpha)$ decreases 
from left to right on either side). Here, $U^\pm_\gamma$ denote the group element
$U_\gamma(t)$ when $t$ lies on the side $c_\pm$ of the wall $\cW(\gamma_1,\gamma_2)$.
Thus, assuming that the $\Omega^+_n(\gamma)$'s are known for all $\gamma\in\tilde\Gamma$,
the $\Omega^-_n(\gamma)$'s can be computed by re-ordering the product on the l.h.s. of 
\eqref{ewallfinU} in the opposite order, using the commutation rule \eqref{KSalg} 
and the Baker-Campbell-Hausdorff (BCH) 
formula, and reading off the exponents $\Omega^-_n(\gamma)$. 

This procedure is vastly simplified by expressing the products in \eqref{ewallfinU}
in terms of the `rational motivic invariants'
\be \label{bOmref}
\bar \Omega_{\rm ref}(\gamma,t,y) 
\equiv \sum_{m|\gamma} 
\frac{(y - y^{-1})}{m (y^m - y^{-m})}
\Omega_{\rm ref}(\gamma/m, t, y^m) \, .
\ee
The relation between $ \Omega_{\rm ref}(\gamma,y)$ and $\bar\Omega_{\rm ref}(\gamma,y)$
is easily inverted by means of the M\"obius formula,
\be \label{bOmrefi}
 \Omega_{\rm ref}(\gamma,t,y) 
= \sum_{m|\gamma} \mu(m)\, 
\frac{(y - y^{-1})}{m (y^m - y^{-m})}
\bar\Omega_{\rm ref}(\gamma/m, t, y^m) \, ,
\ee
where $\mu(d)$ is the M\"obius function (1 if $d$ is a product of an even number 
of distinct primes, $-1$ if $d$ is a product of an odd number of primes, or $0$ otherwise). 
Using the fact that the generators $e_{\ell\gamma}$ commute for all $\ell\in \IZ$, we may rewrite
\eqref{ewallfinref} as a product of factors labelled by coprime $(M,N)$,
\be \label{ewallfinref}
\prod_{\substack{M\geq 0,N\geq 0>0, \\ \gcd(M,N)=1, M/N\downarrow}}
V^+_{M\gamma_1+N\gamma_2}
=
\prod_{\substack{M\geq 0,N\geq 0>0, \\ \gcd(M,N)=1, M/N\uparrow}}
V^-_{M\gamma_1+N\gamma_2}\, ,
\ee
where
\be
\label{Vquant}
V_{\gamma}(t) = \prod_{\ell\geq 1} U_{\ell\gamma}(t) = 
\exp\left(\sum_{N=1}^{\infty} \bar
\Omega_{\rm ref}(N\gamma, t,y)\, 
e_{N\gamma} \right)\ .
\ee
Due to the fact that the algebra \eqref{KSalg} is graded by $\tilde\Gamma$, and that every
factor of $e_{\alpha} $ in the logarithm of $V_{\gamma}$ is multiplied by 
a factor of $\Omega_{\rm ref}(\alpha, y)$ for the same vector $\alpha$, it is clear that the
result of the re-ordering procedure outlined below \eqref{ewallfinU} will produce 
a wall-crossing formula of the form 
\be \label{efirst3}
\bar\Omega_{\rm ref}^-(\gamma,y) -\bar\Omega^+_{\rm ref}
(\gamma,y) =
\sum_{n\geq 2 }\, 
  \sum_{\substack{ \{\alpha_1,\dots, \alpha_n\} \in \tilde\Gamma^n \\
\gamma= \alpha_1+\dots +\alpha_n}}\, 
\frac{g_{\rm ref}(\{\alpha_i\},y)}{\prod m_k!}
 \prod\nolimits_{i=1}^n \bOm_{\rm ref}^+(\alpha_i,y) \ ,
 \ee
 where the second sum runs over unordered sets of  
$n$ charge vectors $\alpha_i=M_i\gamma_1+N_i\gamma_2$ 
such that $\sum_{i=1\dots n} \alpha_i=\gamma=M\gamma_1+N\gamma_2$. 
By the 'root basis' property, only a finite number of terms appear on the 
right-hand side. In \S\ref{sec_prim} below, we shall compute the universal combinatorial
factors $g_{\rm ref}(\{\alpha_i\},y)$ in selected cases.

Before doing so however, we discuss the classical (or numerical) KS wall-crossing
formula, which arises from the motivic formula \eqref{ewallfinref} in the limit $y\to 1$.
In this limit, the Poincar\'e polynomial \eqref{refdeg} reduces to the Euler-Behrend 
characteristic of the moduli space $\cM(\gamma,t)$
\be
\Omega_{\rm ref}(\gamma,t,y) \,\stackrel{y\to 1}{\longrightarrow} \, \chi[\cM(\gamma,t)]\in \IZ\ ,
\ee
while $\bar \Omega_{\rm ref}(\gamma,y)$ reduces to
the `rational DT invariant'
\be
 \label{efirst}
\bar \Omega_{\rm ref}(\gamma,t,y) \, \stackrel{y\to 1}{\longrightarrow}\,  
\bar\Omega(\gamma,t)\equiv \sum_{m|\gamma}m^{-2}\, 
\Omega(\gamma/m,t)\ .
\ee
Moreover, the commutation rule \eqref{KSalg} in the Lie algebra $\cA$
has a smooth limit 
\be
\label{KSalgclas}
[ e_{\gamma} , e_{\gamma'} ] =  (-1)^{\langle \gamma,\gamma'\rangle}\, 
\langle \gamma,\gamma'\rangle\, e_{\gamma+\gamma'}\ ,
\ee
and so does the operator $V_\gamma$ in \eqref{Vquant}.
As a result, the combinatorial coefficients $g_{\rm ref}(\{\alpha_i\},y)$
have a smooth limit as $y\to 1$, and the wall-crossing formula for the 
rational DT invariants is given by the limit of \eqref{efirst3} as 
$y\to 1$, i.e. Eq.  \eqref{jswcf2} with 
\be
g(\{\alpha_i\}) = \lim_{y\to 1} g_{\rm ref}(\{\alpha_i\},y)\ .
\ee

\subsection{Primitive and semi-primitive wall-crossing\label{sec_prim}}

Despite the fact that either side of the KS wall-crossing 
formula \eqref{ewallfinref} involves an infinite number of factors, the 
procedure of re-ordering the product involves only a  finite number of
operations, for the following reason (already hinted at below \eqref{efirst3}):
for any $M,N\geq 0$, 
\be
\cI_{M,N}\equiv \{ \sum_{m>M \, {\rm and/or}\,
n> N} \IR  \cdot e_{m\gamma_1+n\gamma_2} \}
\ee
 is a two-sided  ideal in $\cA$,
and the quotient $\cA_{M,N}=\cA/\cI_{M,N}$ is a finite dimensional algebra. For the purpose
of computing $\Delta\Omega(M\gamma_1+N\gamma_2)$, it is sufficient to project
the relation \eqref{ewallfinref} to $\cA_{M,N}$ and use the truncation of the BCH formula
at order $\min(M,N)$. E.g.  to compute $\Delta\Omega(\gamma_1+\gamma_2)$, it 
suffices to re-order the l.h.s of the identity in $\cA_{1,1}$ 
\ben \label{erel1}
&& \exp(\Omega^+(\gamma_1)e_{\gamma_1})
\exp(\Omega^+(\gamma_1+\gamma_2)
e_{\gamma_1+\gamma_2})
\exp(\Omega^+(\gamma_2)e_{\gamma_2})\nonumber \\
&&\qquad = 
\exp(\Omega^-(\gamma_2)e_{\gamma_2})
\exp(\Omega^-(\gamma_1+\gamma_2)
e_{\gamma_1+\gamma_2})
\exp(\Omega^-(\gamma_1)e_{\gamma_1})\, 
\een
using the truncated BCH formula $e^X\, e^Y = 
e^{X+Y +{1\over 2} [X,Y]}$, and match the result to the r.h.s. In this way, we 
find that the motivic invariants $\Omega_{\rm ref}( \gamma_1,y),
\Omega_{\rm ref}( \gamma_2,y)$ are constant across the wall, 
while 
\be
\label{wcref}
\Delta\Omega_{\rm ref}( \gamma_1+\gamma_2,y) 
= \kappa( \langle \gamma_1,\gamma_2\rangle,y )\, 
\Omega_{\rm ref}(\gamma_1,y)\,
 \Omega_{\rm ref}(\gamma_2,y)\ .
\ee
This relation (or its obvious classical limit at $y=1$) is known
as the `primitive wall crossing formula' \cite{Denef:2007vg,Diaconescu:2007bf,
Dimofte:2009bv}. From \eqref{wcref}, setting $\alpha_1=\gamma_2$ and 
$\alpha_2=\gamma_1$ so that $\langle \alpha_1,\alpha_2\rangle>0$,
we read off the combinatorial factor 
\be
\label{gref2}
g_{\rm ref}(\alpha_1,\alpha_2,y) =  -\kappa( \langle \alpha_1,\alpha_2\rangle,y ) = 
(-1)^{\langle \alpha_1,\alpha_2\rangle+1}\, 
\frac{\sinh ( \nu \langle \alpha_1,\alpha_2\rangle)}{\sinh\nu}\ .
\ee
Up to a sign, this is 
recognized as the character $\Tr(-y)^{2J_3}$ of a representation of $SU(2)$ with 
spin $j=\frac12(|\langle \alpha_1,\alpha_2\rangle|-1)$.

With some more work, one can easily extract the combinatorial coefficients 
$g_{\rm ref}(\{\alpha_i\},y)$ for $n>2$. E.g, for $n=3$ and $\alpha_1,\alpha_2,\alpha_3$
three distinct (non necessarily primitive) elements of $\tilde\Gamma$ 
ordered such that $\alpha_{ij}\equiv \langle\alpha_i, \alpha_j\rangle >0$
for $i<j$, we find
\be \label{gref3}
\begin{split}
g_{\rm ref}(\alpha_1,\alpha_2,\alpha_3, y)
= (-1)^{\alpha_{12}+\alpha_{13}+\alpha_{23}}\,
\frac{\sinh(\nu\alpha_{12}) \sinh(\nu(\alpha_{13}+
\alpha_{23}))}{ \sinh^2\nu}\ 
\end{split}
\ee
when $\alpha_{12}>\alpha_{23}$, or 
\beq 
\label{gref32}
g_{\rm ref}(\alpha_1,\alpha_2,\alpha_3, y)
=(-1)^{\alpha_{12}+\alpha_{13}+\alpha_{23}}\nn\,
\frac
{\sinh(\nu\alpha_{23}) \sinh(\nu(\alpha_{12}+
\alpha_{13}))}{\sinh^2\nu}\ 
\eeq
when $\alpha_{12}<\alpha_{23}$. The result for $n=4$ can be found in  \cite{Manschot:2010qz}.

While the amount of work necessary to extract the combinatorial factors quickly
grows with $(M,N)$, for fixed (small) $M$ it is possible to compute all the jumps
for $\gamma\to M\gamma_1+N\gamma_2$ at once using the Hadamard
lemma $\log(e^X\, Y\, e^{-X})=\sum_{n\geq 0} Ad_X^n\cdot Y)/n!$, where
$Ad_X\cdot Y\equiv [X,Y]$. Defining
\be
Z^\pm(M,q,y)= \sum_{N=0}^{\infty} 
\Omega_{\rm ref}^\pm(M,N,y)\, q^N\ ,\quad 
\Omega_{\rm ref}^\pm(M,N,y)\equiv \Omega_{\rm ref}^\pm(M\gamma_1+N\gamma_2,y)\ ,
\ee
we find, for $M=1$ \cite{AndriyashMoore,Manschot:2010qz}, 
the 'semi-primitive' wall crossing formula \cite{Denef:2007vg}
\be
\label{Z1pmy}
Z^-(1,q,y) = Z^+(1,q,y)\,Z_{\rm halo}(\gamma_1,q,y) 
\ee
where
\be 
\label{ezhalo}
Z_{\rm halo}(\gamma_1,q,y) \equiv 
\exp\left(
\sum_{\ell=1}^{\infty} 
\kappa( \ell \langle \gamma_1, \gamma_2 \rangle, y)
\,  \bOm_{\rm ref}(\ell\gamma_2,y)\, q^\ell \right)\ .
\ee
The reason for the subscript `halo' will become apparent in \S\ref{sec_stat}. Re-expressed
in terms of the integer motivic invariants, this can be written as an infinite 
product \cite{Dimofte:2009tm}
\be
\label{Zhalomotprod}
Z_{\rm halo}(\gamma_1,q,y) = \prod_{\substack{k\geq 1, n\in\IZ\\
1\leq j\leq k |\gamma_{12}| }}
\left( 1- (-1)^{k |\gamma_{12}|} q^k y^{n+2j
-1-k |\gamma_{12}|}\right)^{ (-1)^n\, 
\Omega_{n}(k\gamma_2)}
\ee
Generalizations of \eqref{Z1pmy} for $M=2,3$ can be found in \cite{Manschot:2010qz}.

\subsection{Exact wall crossing}

Finally, we discuss some examples where the re-ordering of the product in 
\eqref{ewallfinref} can be performed in the full untruncated algebra $\cA$. 
Suppose that in the chamber $c_+$, the only non-vanishing DT invariants
are $\Omega^+(\gamma_1)$ and $\Omega^+(\gamma_2)$. If $\gamma_{12}=-1$,
the result of the re-ordering gives\footnote{In \eqref{pentawcf} and \eqref{sw0wcf},
$\Omega=1$ for each factor, except for the factor $U_{(2,0)}$ in the middle
of \eqref{sw0wcf}, for which $\Omega=-2$.}
\be
\label{pentawcf}
U_{\gamma_2}\, U_{\gamma_1} = U_{\gamma_1}\, U_{\gamma_1+\gamma_2}\, U_{\gamma_2}\ ,
\qquad \gamma_{12}=-1\ ,
\ee
which follows from the pentagonal identity for the quantum dilogarithm function.
If instead $\gamma_{12}=-2$, one arrives at \cite{ks}
\be
\label{sw0wcf}
U_{(2,-1)} \cdot U_{(0,1)} = U_{(0,1)}\cdot U_{(2,1)}\cdot U_{(4,1)}\dots U_{(2,0)}\dots 
U_{(3,-1)} \cdot  U_{(2,-1)} U_{(1,-1)}\ ,  
\ee
where we denoted $\gamma_2=(0,1)$, $\gamma_1=(2,-1)$  to match
the usual basis of electromagnetic charges in Seiberg-Witten theory with $G=SU(2)$
and no flavors \cite{Seiberg:1994aj}. As first noted by Denef, 
Eq. \eqref{sw0wcf} then embodies the BPS 
spectrum of this gauge theory on the two sides of the curve of marginal stability 
$\Im(a/a_D)=0$, see Fig. \ref{fig_sw0}. Analogues of \eqref{sw0wcf} for $SU(2)$
gauge theories with $0<N_f<4$ flavors can be found in \cite{Gaiotto:2008cd,Dimofte:2009tm}.
More general identities of this type can be derived using Y-systems and 
cluster algebra techniques,
see e.g. \cite{Gliozzi:1994cs, MR1362962,Nakanishi:2010}.

\section{The Joyce-Song wall-crossing formula \label{sec_js}}

In this section, we briefly review the Joyce-Song wall-crossing formula, 
which was derived in the context of the Abelian category of coherent 
sheaves on a Calabi-Yau three-fold $\cX$ \cite{Joyce:2008pc}. Unlike
the KS formula, the JS formula only applies to the jump of the classical
(or numerical) Donaldson-Thomas invariants. Moreover, it gives a fully
explicit formula for the combinatorial factors $g(\{\alpha_i\})$ 
appearing in \eqref{jswcf2}. The price to pay is that the JS formula 
is computationally less efficient, as it involves sums over many terms
with large denominators and large cancellations. 

To state the JS formula, 
we first  introduce $S$, $U$ and $\cL$ factors,
which are functions of an ordered list of charge vectors $\alpha_i\in\tilde\Gamma,
i=1\dots n$:
\begin{itemize}
\item 
We define $S(\alpha_1,\ldots,\alpha_n)\in \{0,\pm 1\}$ as follows.
If $n=1$, set $S(\alpha_1)=1$. 
If $n>1$ and, for every $i=1\dots n-1$, either 
\beq
(a) &&\quad \langle \alpha_i,\alpha_{i+1}\rangle \leq 0 
\quad \mbox{and}\quad
\langle \alpha_1+\cdots+\alpha_i,
\alpha_{i+1}+\cdots+\alpha_n \rangle < 0 , \quad\mbox{or}\, 
\nn \\
(b) && \quad  \langle \alpha_i,\alpha_{i+1} \rangle > 0 
\quad \mbox{and}\quad
 \langle \alpha_1+\cdots+\alpha_i, \alpha_{i+1}+\cdots+\alpha_n 
 \rangle \geq 0 \ ,
 \eeq
let $S(\alpha_1,\ldots,\alpha_n)=(-1)^r$, where $r$ is the 
number of times option (a) is realized; otherwise, 
$S(\alpha_1,\ldots,\alpha_n)=0$. 

\item To define the $U$ factor (not to be confused with the operator $U$ of the previous 
section~!), consider all ordered partitions of the $n$ vectors 
$\alpha_i$ into $1\leq m\leq n$ packets $\{\alpha_{a_{j-1}+1},\cdots,\alpha_{a_j}\}$, $j=1\dots m$,
with $0=a_0<a_1<\cdots<a_m=n$, such that all vectors within each packet are collinear. Let 
\be
\beta_j=\alpha_{a_{j-1}+1}+\cdots+\alpha_{a_j}, \qquad j=1\dots m
\ee
be the sum of the charge vectors in each packet.
Next, consider all ordered partitions of the $m$ vectors $\beta_j$
into $1\leq l\leq m$ packets
$\{\beta_{b_{k-1}+1},\cdots,\beta_{b_k}\}$, with
$0=b_0<b_1<\cdots<b_l=m$, $k=1\dots l$, such that the total charge
vectors $\delta_k=\beta_{b_{k-1}+1}+\cdots+\beta_{b_k}, k=1\dots l$ are all collinear.
Define the $U$-factor as the sum
\be
\begin{split} 
U(\alpha_1,\ldots,\alpha_n) &\equiv  \sum_l 
\frac{(-1)^{l-1}}{l}\cdot\prod\nolimits_{k=1}^l \\
& \prod_{j=1}^m\frac{1}{(a_j-a_{j-1})!}\,\, S(\beta_{b_{k-1}+1},
\beta_{b_{k-1}+2},\ldots,\beta_{b_k})\ .
\end{split}
\ee
over all partitions of $\alpha_i$ and  $\beta_j$
satisfying the conditions above. If none of the $\alpha_i$ are parallel, $S=U$. 
Contributions with $l>1$ arise only when $\{\alpha_i\}$
can be split into two (or more) packets with the same total charge, e.g.
\be
U[ \gamma_1,\gamma_2,\gamma_1,\gamma_2]= 
   S[\gamma_1,\gamma_2,\gamma_1,\gamma_2]
   -\frac{1}{2} S[\gamma_1,\gamma_2]^2 = 1-\frac{1}{2}(-1)^2=\frac12 
\ee

\item Finally, departing from the notations in \cite{Joyce:2008pc}, define 
the $\cL$ factor  by
\be
\cL(\alpha_1,\dots,\alpha_n) = 
\sum_{\rm trees}\,\
\prod\limits_{\rm edges (i,j)}  \langle \alpha_{i}, \alpha_{j} \rangle
\ee
where the sum runs over all labelled trees with $n$ vertices
labelled $ \{1,\dots,n \}$, with edges oriented from $i$ to $j$ if $i<j$.
There are $n^{n-2}$  labelled trees with $n$-vertices, which can
be labelled by their 
Pr\"ufer code, an arbitrary sequence of $n-2$ numbers
in $\{1,\dots n\}$.

\end{itemize}

With these definitions, the result of \cite{Joyce:2008pc} can be stated as an explicit 
formula for the combinatorial factors $g(\{\alpha_i\})$ 
appearing in \eqref{jswcf2}:
\be 
\label{egfromjs}
%\begin{split}
g(\{\alpha_i\})\,  = %&
\frac{1}{2^{n-1}}
\, (-1)^{n-1+\sum_{i<j} \langle \alpha_i,\alpha_j\rangle}
\sum_{\sigma\in\Sigma_n} \\
%& 
\cL\left(\alpha_{\sigma(1)},\dots \alpha_{\sigma(n)}\right)\, 
U\left(\alpha_{\sigma(1)},\dots \alpha_{\sigma(n)}\right)\ .
%\end{split}
\ee

As an illustration, we now use the JS formula to derive the combinatorial
factors $g(\{\alpha_i\})$ for $n=2,3$. For $n=2$ (assuming as before that 
$\gamma_{12}<0$), the $U,S,\cL$ factors are given in the following table 
\be
\label{tg12}
\begin{array}{|c|c|c|c|}
\hline
\sigma(12) & S & U & \cL \\
\hline
12 & \text{a} & -1 & \gamma_{12}\\
21 & \text{b} & 1 & -\gamma_{12}\\
\hline
\end{array}
\ee
The JS formula \eqref{egfromjs} then leads to
\be
g(\gamma_1,\gamma_2)=(-1)^{\gamma_{12}}\, \gamma_{12}\, 
\Omega(\gamma_1)\, \Omega(\gamma_2)\ ,\qquad
%\gamma_{12}\equiv \langle \gamma_1, \gamma_2 \rangle\ ,
\ee
in agreement with the classical limit of \eqref{gref2}. 

For $n=3$, assuming as before that $\alpha_1,\alpha_2,\alpha_3$ are
three distinct (non necessarily primitive) elements of $\tilde\Gamma$ 
ordered such that $\alpha_{ij}\equiv \langle\alpha_i, \alpha_j\rangle >0$
for $i<j$ and moreover that $\alpha_{12}>\alpha_{23}$, we find 
that the $S,U,\cL$ factors are given by 
\be
\label{tg123}
\begin{array}{|c|c|c|c|}
\hline
\sigma(123) & S & U & \cL \\
\hline
123 & \text{bb} & 1 & \alpha_{12}\alpha_{13}+\alpha_{13}\alpha_{23}+\alpha_{12}\alpha_{23}\\
132 & \text{b-} & 0 & \alpha_{12}\alpha_{13}-\alpha_{13}\alpha_{23}-\alpha_{12}\alpha_{23}\\
213 &\text{ab} & -1 & -\alpha_{12}\alpha_{23}+\alpha_{13}\alpha_{23}-\alpha_{12}\alpha_{13}\\
231 &  \text{-a} & 0 & \alpha_{12}\alpha_{13}-\alpha_{13}\alpha_{23}-\alpha_{12}\alpha_{23}\\
312 & \text{ab} & -1 & \alpha_{13}\alpha_{23}-\alpha_{12}\alpha_{23}-\alpha_{13}\alpha_{12} \\
321 & \text{aa} & 1 & \alpha_{13}\alpha_{23}+\alpha_{12}\alpha_{13}+\alpha_{12}\alpha_{23} \\
\hline
\end{array}
\ee
The JS formula \eqref{egfromjs} then leads to
\be
g(\{\alpha_1,\alpha_2,\alpha_3\}) = (-1)^{\alpha_{12}+\alpha_{23}
 +\alpha_{13}}\, \alpha_{12} \left( \alpha_{13} 
+ \alpha_{23}\right) \ ,
\ee
in agreement with the classical limit of \eqref{gref3}. 
The computation for $n=4$ can be found in  \cite{Manschot:2010qz}, and matches
the result from the KS formula.

\section{Wall-crossing from multi-centered quantum black holes \label{sec_wcbh}}

In this section, we give a new physical derivation of the wall-crossing formula 
\eqref{jswcf2} (in particular, a new formula for the combinatorial factors 
$g(\{\alpha_i,y\})$) based on the quantum mechanics of multi-centered black 
hole configuration. Before starting, it should be noted that the KS wall-crossing
formula (and to a lesser extent, the JS formula) has already been derived
or interpreted in various physical settings \cite{Diaconescu:2007bf,Gaiotto:2008cd,
Alexandrov:2008gh,Jafferis:2008uf,Cecotti:2009uf,Cecotti:2010fi,Gaiotto:2010be,
Andriyash:2010qv,Manschot:2009ia}. Our derivation is arguably more elementary,
as it relies only on the supersymmetric quantum mechanics of point particles 
interacting by Coulomb and Lorentz-type  forces. The down-side  is that it does not make the 
algebra $\cA$ manifest and, admittedly, relies on some plausible but
not rigorously proven assumptions.

\subsection{From Bose-Fermi to Boltzmann statistics \label{sec_stat}}

To motivate our approach, let us return to the semi-primitive wall-crossing formula
\eqref{Z1pmy} and for simplicity, concentrate on the classical limit $y=1$. 
Substituting \eqref{efirst} in \eqref{ezhalo}, one may rewrite \eqref{Z1pmy}
as
\be
\label{Z1pm1}
\frac{\sum\nolimits_{N\geq 0} \Omega^-(1,N)\, q^N}
{\sum\nolimits_{N\geq 0} \Omega^+(1,N)\, q^N}
= \prod_{k>0} 
\left( 1- (-1)^{k \gamma_{12}} q^k\right)^{ k \, 
|\gamma_{12}|\, 
\, \Omega^+(k\gamma_2)}\ .
\ee
E.g.  for $\gamma\mapsto \gamma_1+2\gamma_2$, we find
\be
\label{dom12}
\begin{split}
\Delta\Om(1,2)=&
2 {\gamma_{12}}\, \Om^+(1,0) \, \Om^+(0,2) 
+ (-1)^{{\gamma_{12}}}\,  {\gamma_{12}}\, \Om^+(1,1)   \Om^+(0,1)
 \\&
+\Om^+(1,0)\, \left[  \frac12 {\gamma_{12} \, \Om^+(0,1) \left( \gamma_{12} \Om^+(0,1) +1 \right) } \right]\ . 
\end{split}
\ee
The two contributions on the first line can be interpreted as the index of two-centered
black hole solutions, carrying charges $\alpha_1=\gamma_1$ and $\alpha_2=2\gamma_2$ 
for the first term, or $\alpha_1=\gamma_1+\gamma_2$ and $\alpha_2=\gamma_2$ for the 
second term. Indeed, for such two-centered solutions, the distance is 
fixed to~ \cite{Denef:2000nb} 
\be
r_{12} = \frac12 \frac{\langle \alpha_1, \alpha_2\rangle \, |Z(\alpha_1)+ Z(\alpha_2)|}{\Im[ Z(\alpha_1)\bar Z(\alpha_2)] }\ ,
\ee
which is positive on the side $c_-$ of the wall only; the unit vector $\vec r_{12}/r_{12}$ can
be chosen arbitrarily on the unit two-sphere. Such configurations carry angular momentum
$\vec J=\frac12(\alpha_{12}-1)\vec r_{12}/r_{12}$ (similar to the angular momentum carried by 
an electron in a magnetic monopole background), and therefore have $|\alpha_{12}|$
possible configurational states, with index 
$g(\alpha_1,\alpha_2)=(-1)^{\alpha_{12}} \alpha_{12}$. 
Since the distance $r_{12}$ diverges in the vicinity of the wall, 
the internal degrees of freedom are decoupled from the configurational degrees of freedom,
and the total index is the product $g(\alpha_1,\alpha_2)\Omega^-(\alpha_1)
\Omega^-(\alpha_2)$, consistently with \eqref{dom12}.

The contribution on the second line of \eqref{dom12} is more interesting.
Letting $d=\gamma_{12} \Om^+(0,1)$, the term in bracket is recognized 
as the index of the symmetric  (or, when $d<0$, antisymmetric) part of 
the tensor product $\cH_1\otimes \cH_1$, where $\cH_1$ is the 
space of quantum states accessible to  one particle of charge $\gamma_2$ in the field
of the particle of charge $\gamma_1$. In other words, the second line corresponds
to a configuration of two identical centers of charge $\gamma_2$ orbiting around 
a center of charge $\gamma_1$, with Bose statistics when $d>0$, or Fermi statistics
when $d<0$.  More generally, \eqref{Z1pm1} can be interpreted as the contribution
from halos of particles of charge $k\gamma_2$ orbiting on a fixed shell around a center of charge
$\gamma_1+k_0\gamma_2$, and obeying Bose or Fermi statistics, depending on the sign
of $\langle \gamma_1+k_0 \gamma_2, k \gamma_2\rangle\, \Omega^+(k\gamma_2)$.

In contrast,  in terms of the rational DT invariants 
the semi-primitive wall-crossing formula \eqref{Z1pmy} reads
\be
\label{Z1pm1b}
\frac{\sum\nolimits_{N\geq 0} \Omega^-(1,N)\, q^N}
{\sum\nolimits_{N\geq 0} \Omega^+(1,N)\, q^N}
= \exp\left(
\sum_{\ell=1}^{\infty} 
(-1)^{\ell  \langle \gamma_1, \gamma_2 \rangle}\, \ell  \langle \gamma_1, \gamma_2 \rangle
\,  \bOm(\ell\gamma_2,y)\, q^\ell \right)\ ,
\ee
so that, e.g. for $\gamma\mapsto \gamma_1+2\gamma_2$, 
\be
\label{dom12b}
%\begin{split}
\Delta\Om(1,2)=%&
2 {\gamma_{12}}\, \Om^+(1,0) \, \bOm^+(0,2) 
+ (-1)^{{\gamma_{12}}}\,  {\gamma_{12}}\, \Om^+(1,1)   \Om^+(0,1)
% \\&
+\Om^+(1,0)\, \left[  \frac12 \left(  \gamma_{12} \, \Om^+(0,1) \right)^2  \right]\ ,
%\end{split}
\ee
where we combined $\Omega(0,2)$ and $-\frac14\bar\Omega(0,1)$ in \eqref{dom12}
into $\bOm(0,2)$. Unlike \eqref{dom12}, the last term in \eqref{dom12b} is of the form 
$\frac12 d^2$, which would be the result if the two particles of charge $(0,1)$ were 
distinguishable and obeyed Boltzmann statistics. More generally, \eqref{Z1pm1b}
can be interpreted as contributions of the same halo of particles with 
charge $k\gamma_2$ described above, but now satisfying Boltzmann statistics. 
While \eqref{dom12b} is hardly shorter than \eqref{dom12}, the reader can easily
convince him/herself of the power of this simplification by computing $\Delta\Omega(1,N)$
for higher $N$.

The lesson to take from this discussion is that, rather than computing the variation
of $\Omega(\gamma,t)$ across the wall $\cW(\gamma_1,\gamma_2)$, it is advantageous 
to compute instead the variation of the rational invariants $\bar\Omega(\gamma,t)$ defined
in \eqref{efirst}, and apply the following recipe: treat the centers as distinguishable 
pointlike particles and compute their configurational index. 
Whenever $m$ centers carry the same charge $\alpha$, divide the configuration index 
by a Bolztmann-Gibbs
factor $1/m!$. Finally, multiply the configurational 
index  by the effective (rational) index $\bar\Omega(\alpha,t)$ carried by each of the centers. 
The result of this recipe is the formula 
\eqref{jswcf2}, where $g(\{\alpha_i\})$ is identified as the  configurational index of the
quantum mechanics of $n$ distinguishable particles  interacting via Coulomb and 
Lorentz type forces
(which we discuss in detail in the next subsection). Although we motivated this recipe
by inspecting the classical semi-primitive formula, it in fact holds  in full generality
and applies to  the refined (or motivic)
index as well, see \cite{Manschot:2010qz} for more details.

\subsection{The phase space of multi-centered BPS black holes \label{sec_sugra}}

Let us now review some relevant
properties of supersymmetric multi-centered black hole 
solutions in $\cN=2$ supergravity (a similar analysis for multi-centered
dyon solution in the low energy limit of $\cN=2$ gauge theories can be
found in \cite{Denef:1998sv,Lee:2011ph}).  Such solutions
fall into the stationary metric ansatz 
\be
\de s^2= - 
e^{2U} \, (\de t+\cA )^2 + e^{-2U} \de \vec r^2 
\ee
where the scale function $U$, the Kaluza-Klein 
one-form $\cA$ and the vector multiplet scalars $z^a,
a=1\dots n_v$ depend on the coordinate $\vec r$ on 
$\IR^3$. 

For $n$ centers located at $\vec r_1,\dots, 
\vec r_n$, carrying electromagnetic charges $\alpha_1,\dots ,
\alpha_n\in\Gamma$ with total charge 
$\gamma=\alpha_1+\dots+\alpha_n$,
the values of the vector multiplet
scalars $t$ and of the scale factor $U$
are obtained by solving  \cite{Denef:2000nb}.
\be
\label{eqatt}
-2\, e^{-U(\vec r)} \Im\left[ e^{-\I\phi} Y(t(\vec r)) \right]=  \beta +
\sum_{i=1}^n {\alpha_i\over |\vec r - \vec r_i|},
\qquad \phi=\arg Z(\gamma,t_\infty),
\ee
where $Y(t)=-e^{\cK/2}(X^\Lambda(t),F_\Lambda(t))$ is the
symplectic section afforded by the special geometry of the
vector multiplet moduli space, such that $Z(\gamma,t)=\langle \gamma, Y(t_\infty)\rangle$.
The constant vector $\beta$
on the right-hand side of \eqref{eqatt} is determined in 
terms of  the asymptotic values of the moduli at infinity $t_\infty$
by 
\be \label{edefbeta}
\beta = -2 \, {\rm Im} \left[e^{-\I\phi}\, Y(t_\infty) \right]\, .
\ee
In particular, it follows from \eqref{eqatt} that the scale factor $U$ 
is given by evaluating the Bekenstein-Hawking entropy function 
$S(\gamma)$ on the harmonic function appearing on the right-hand side
of \eqref{eqatt} \cite{Bates:2003vx},
\be
\label{esk2aa}
e^{-2U(\vec r)}  = \frac{1}{\pi} S\left( \beta +\sum_{i=1}^n {\alpha_i\over |\vec r - \vec r_i|}
\right)\ .
\ee
Most importantly for our purposes, the locations $\vec r_i$ are subject to the 
condition of mechanical equilibrium under the Coulomb, Lorentz, Newton and 
scalar exchange forces
(also known as integrability equations) \cite{Denef:2000nb}
\be \label{denef3d}
\sum_{j=1\atop j\ne i}^n \frac{\alpha_{ij}}{r_{ij}}
=  c_i\, ,
\ee
where $r_{ij} = |\vec r_i - \vec r_j|$, $\alpha_{ij}
\equiv \langle \alpha_i, \alpha_j\rangle$, 
and the real constants 
\be
\label{eci}
c_i \equiv 2 \, {\rm Im}\, [e^{-\I\phi} Z(\alpha_i,t_\infty)]
\ee
depend on the the asymptotic values of the moduli. Since $\phi=\arg Z(\gamma,t_\infty)$,
the constants $c_i$ satisfy $\sum_{i=1}^n c_i=0$.
The conditions  \eqref{denef3d} 
guarantee the existence of a Kaluza-Klein connection
$\cA$ such that the above configuration is a supersymmetric solution
of the equations of motion. In order for the solution to be physical however,
one must also
require that the scale factor be everywhere 
positive
\be \label{eregu}
S\left( \beta +
\sum_{i=1}^n {\alpha_i\over |\vec r - \vec r_i|}
\right) > 0\, , \qquad \forall \ \vec r\in \IR^3\, ,
\ee
where $\vec r_i$ is the location of the $i$-th
center. For the configurations relevant to the wall-crossing problem, 
this condition appears to be automatically satisfied.

\begin{figure}
\centerline{\includegraphics[height=8cm]{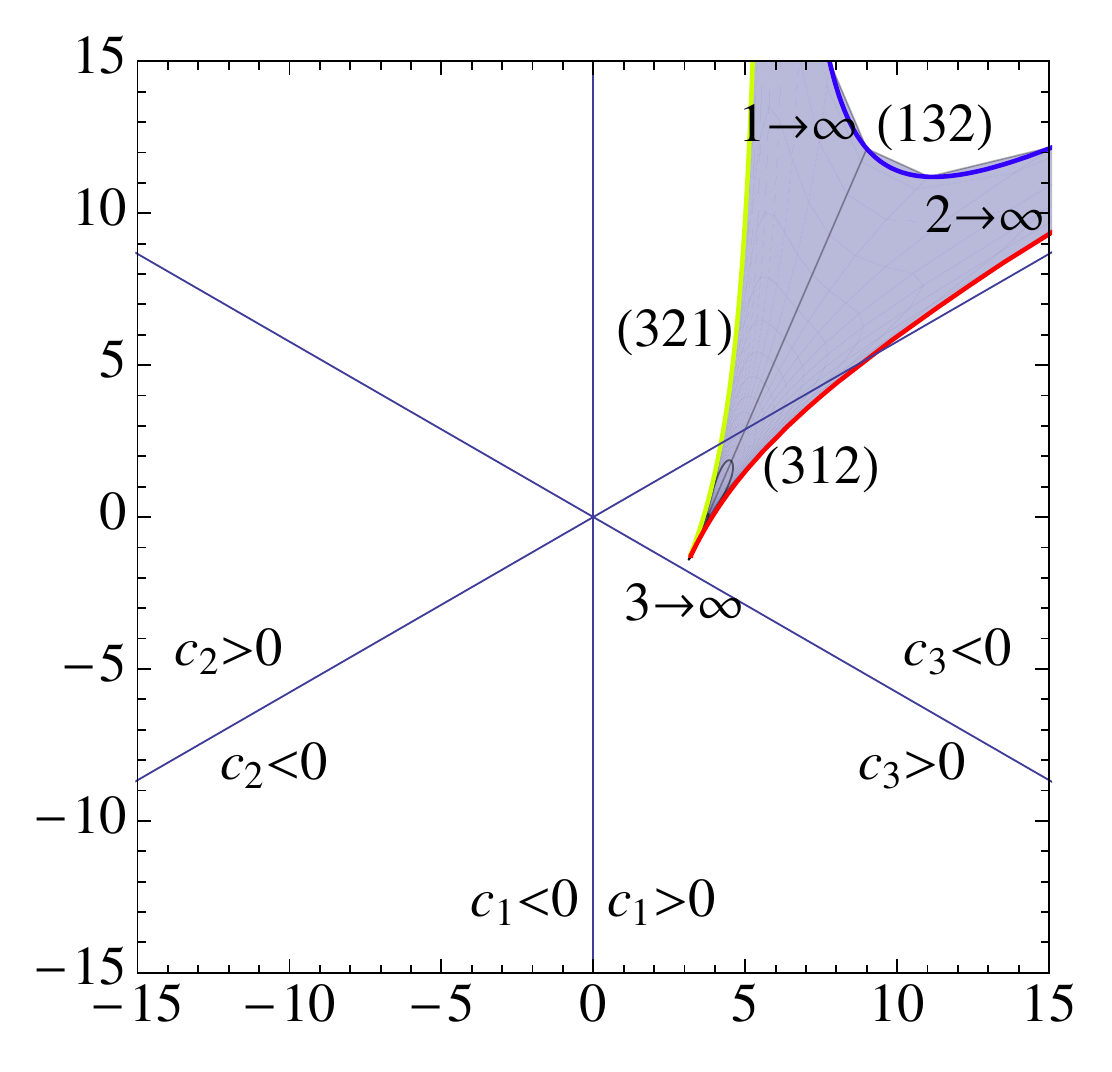} }
\caption{Phase structure of the  moduli space $\cM_n$ of 3-centered solutions as 
a function of $c_i$, for fixed charges such that $\alpha_{12}>0,\alpha_{23}>0,\alpha_{13}>0$,
$\alpha_{12}<\alpha_{23}$. 
The shaded area represent the values of $c_i$ in the two-dimensional section 
$c_1+c_2+c_3=0$ which are spanned as the location of the 3rd center is varied, 
keeping the centers 1 and 2 fixed. Conversely, if the values of $c_i$ is fixed, the range 
of distances between the centers 1 and 2 can be read off by intersecting the shaded
area with a radial line which joins $c_i$ to the origin. Thus, for this choice of charges, 
3-centered solutions only exist in the region $c_1>0, c_3<0$. Inside this region, the range
of $r_{12}$ is bounded from below and from above, except on the wall of marginal stability $c_2=0$. The boundaries
of the shaded region correspond to collinear solutions whose order is indicated. As
the wall is crossed, the topology of the collinear solutions changes from 
$(321),(132)$ to $(321),(312)$.
\label{fig_cplane}}
\end{figure}

Now, let $\cM_{n}(\{\alpha_{ij}\}; \{c_i\})$ be the space of
of solutions $\{\vec r_1,\dots \vec r_n\}$
to the equilibrium conditions \eqref{denef3d}, 
modulo overall translations of the centers. 
$\cM_n$ is a (possibly disconnected)
$2n-2$-dimensional submanifold of $\IR^{3n-3}\backslash \Delta$,
where $\Delta$ is the locus in $\IR^{3n-3}$ where two or more 
of the centers $\vec r_i$ coincide. $\IR^{3n-3}\backslash \Delta$
is equipped with the closed two-form
\be \label{edefomega}
\omega =\frac14 \sum_{i<j} \alpha_{ij}\, 
\frac{\epsilon^{abc} \de  r^a_{ij} \wedge  \de r^b_{ij}  \, r^c_{ij} 
}{|r_{ij}|^3}\ .
\ee 
For generic values of $c_i$, the restriction of $\omega$ to $\cM_n$ is 
non-degenerate and endows $\cM_n$ with a symplectic structure  \cite{deBoer:2008zn}. 
Moreover,  the symplectic form $\omega$ is invariant under $SO(3)$ rotations. The 
moment map associated to infinitesimal rotations is the  angular momentum
\be \label{ejexp}
\vec J= \frac12 \sum_{i<j} \alpha_{ij}\, 
\frac{\vec r_{ij}}{|r_{ij}|} \ .
\ee 
Away from walls of marginal stability, the distances $r_{ij}$ are bounded from above.
If it is possible to order the $\alpha_i$'s such that $\langle \alpha_i,\alpha_j\rangle\geq 0$
whenever $i\leq j$, as it is the case when  the $\alpha_i$'s lie in a two-dimensional 
cone $\tilde\Gamma$, the distances $r_{ij}$ are also bounded from below by 
a non-zero $r_{\rm min}>0$, and the space $\cM_n$ is therefore compact. E.g. for two centers, 
$\cM_2=S^2$ equipped with $\omega=\frac12 \alpha_{12} \sin\theta\, \de\theta\de\phi$
for $\sign(\alpha_{12})=\sign(c_1)$, and zero otherwise. A representative of the 
 phase structure of $\cM_3$ is illustrated in 
Figure \ref{fig_cplane}.

\subsection{Equivariant index and localization}

The symplectic space $\cM_n\equiv \cM_{n}(\{\alpha_{ij}\}; \{c_i\})$  defines the classical 
phase space of the configurational degrees of freedom of $n$-centered BPS black hole 
solutions. Since such configurations are stationary, the 
Hamiltonian vanishes and all points in $\cM_n$ are degenerate in energy. 
Quantum mechanically, the Hilbert space consists of sections of $\cH=S\otimes \cL$, where
$S=S_+\oplus S_-$ is the total spin bundle\footnote{We assume that $\cM_n$ is spin, see
\cite{MPS2-in-progress} for a discussion of this issue.} over $\cM_n$ and 
$\cL$ is a complex line bundle over $\cM_n$ with first Chern class $\omega$.
BPS states correspond to zero-modes of the Dirac operator $D$ on $\cM_n$,
and are analogous to states in  the lowest Landau level
for an electron immersed in a magnetic flux $\omega$. The Dirac
operator decomposes as $D=D_++ D_-$ where $D_+$
maps $S_+\otimes\cL$ to $S_-\otimes\cL$ and vice-versa. 
The action of $SO(3)$ on $\cM_n$ lifts to an action of $SU(2)$
on $S_\pm\otimes\cL$, and the refined index is then
\be
g_{\rm ref}(\{\alpha_i\},y)  = \Tr_{{\rm Ker}D_+}  (-y)^{2J_3} +
 \Tr_{{\rm Ker}D_-}  (-y)^{2J_3} 
\ee
where $J_3$ is the operator representing the rotations along the $z$ axis. Assuming that 
${\rm Ker} D_-=0$, it follows that the refined index is equal to the equivariant 
index of the  Dirac operator $D$,
\be
\label{equivi}
g_{\rm ref}(\{\alpha_i\},y)  = \Tr_{{\rm Ker}D_+}  (-y)^{2J_3} -
 \Tr_{{\rm Ker}D_-}  (-y)^{2J_3} 
\ee
The assumption that ${\rm Ker} D_-=0$ can be proven when $\cM_n$ is \kahler
(which is the case when $n=2,3$). We do not know how to prove it in general, but it
is supported by the fact that it leads to results in agreement with the KS and
JS formulae. In \S\ref{sec_conclu}, we speculate that this assumption may be  
unnecessary if one were to compute the jump of the  protected spin character
in the context of  $\cN=2$ SYM theories.

Now, by the Atiyah-Bott  Lefschetz fixed point 
formula \cite{MR0212836,MR0232406,MR805808,MR1215720},
 the equivariant index 
localizes to the fixed points of the action of $J_3$ on $\cM_n$.
Clearly, those correspond to solutions where all centers lie 
along the $z$ axis, 
and satisfy the one-dimensional equilibrium conditions
\be \label{denef1d}
\sum_{j=1\atop j\ne i}^n \frac{\alpha_{ij}}
{|z_{j} - z_{i}|} \, 
= c_{i}\ ,\qquad \sum_{i=1}^n z_i=0\ ,
\ee
where the last  equation fixes the
translational zero-mode. 
For any  $\sigma\in \cS_n$, where $\cS_n$ denotes the set of permutations of $\{1,\dots , n\}$, 
we denote by $\cC(\sigma)$ 
the set of solutions to \eqref{denef1d} such that $z_{\sigma(i)}<z_{\sigma(j)}$ if $i<j$. 
The set $\cC(\sigma)$ corresponds to the subset of the critical points 
of  the `superpotential' 
\be
\label{defhatW}
W(\lambda, \{z_i\}) = -\sum_{i<j} \alpha_{ij}\,  {\rm sign}(z_j-z_i)\, 
\ln| z_j - z_i| - \sum_i (c_i - \lambda/n)  z_i 
\ee
which are ordered according to the permutation $\sigma$. 
In the vicinity of a fixed point $p\in \cC(\sigma)$, the angular momentum $J_3$
and the symplectic form $\omega$ take the form
\be \label{ej3omega}
J_3 = {1\over 2} \sum_{i<j} \alpha_{\sigma(i)\sigma(j)}
- {1\over 4} M_{ij}(p)\, (x_i x_j + y_i y_j) + \cdots,
\qquad \omega = \frac12 
M_{ij}(p) \, \de x_i\wedge \de y_j +\cdots \, ,
\ee
where $M_{ij}$ is the Hessian matrix of $W(\lambda, \{z_i\})$ with 
respect to $z_1,\dots z_n$, and $(x_i,y_i)$ are coordinates in the plane
transverse to the $z$-axis at the center $i$, subject to the condition 
$\sum_i x_i=\sum y_i=0$. Except for an overall translational zero-mode,
the matrix $M_{ij}$ is non-degenerate, and the critical points are isolated, so $\cC(\sigma)$
is a finite set (possibly empty).

The Lefschetz fixed point formula of \cite{MR805808} yields an explicit
formula for the refined index 
\be \label{eclassq} 
g_{\rm ref} (\{\alpha_i\},y) = \frac{(-1)^{\sum_{i<j} \alpha_{ij} +n-1}}
{(y-1/y)^{n-1}} \, \sum_{\sigma\in \cS_n} \,  
s(\sigma)\, 
y^{\sum_{i<j} \alpha_{\sigma(i)\sigma(j)}}%\right]
\, ,
\ee
where $s(\sigma)$ counts (with sign) the number of solutions to \eqref{denef1d}
ordered according to the permutation $\sigma$, 
\be \label{esigneq}
s(\sigma) =- \sum_{p\in \cC(\sigma)}\, \sign \det \hat M\, ,
\ee
where $\hat M$ is the Hessian of $W$ as a function of the $n+1$ variables
$\lambda, z_1,\dots z_n$,  evaluated at the given solution
of \eqref{denef1d}. The factor $(y-1/y)^{n-1}$ in \eqref{eclassq}
originates form the equivariant $\hat A$-genus in the Lefschetz 
fixed point formula.
It is convenient to let the sum in \eqref{eclassq}
run over all permutations and set  $s(\sigma)=0$ when there are no solutions to
\eqref{denef1d} in the order specified by $\sigma$. For reasons that will
become clear in \S\ref{sec_qui}, we refer to \eqref{eclassq} as the `Coulomb branch
wall-crossing formula'.

The formula \eqref{eclassq} is fully explicit, yet it depends on our ability
to find solutions of the one-dimensional problem \eqref{denef1d}. While this
can be done numerically  (approximate solutions are sufficient since the answer
depends only on the order $\sigma$ and the sign of $\det M$), it would be useful
to have a general criterium for determining when solutions exist, and if so
to compute their Morse index. The answer to these questions 
is suggested by an alternative 
approach based on quivers (see \S\ref{sec_conclu}).

At this point, we can check whether \eqref{eclassq} agrees with the answer
of the KS or JS formulae. For $n=2$, we find two fixed points with permutations
$\sigma(12)=12$ and $21$, leading to  
\be
g(\alpha_1,\alpha_2;y) = (-1)^{\alpha_{12}}  \frac{
\sinh(\nu \alpha_{12})}{\sinh\nu}\ ,
\ee
in agreement with \eqref{gref2}.
For $n=3$ (and for the same choice of $\alpha_i$ as above \eqref{gref3}), 
we find 4 possible orderings, with the following value of $s(\sigma)$,
\be \label{e3perm}
\{1,2,3;+\}, \{2,1,3;-\}, \{3,1,2;-\}, \{3,2,1;+\}\, ,
\ee
leading to 
\ben \label{e3res}
g_{\rm ref}(\alpha_1,\alpha_2,\alpha_3,y)
&=& (-1)^{\alpha_{12}+\alpha_{23}+\alpha_{13}}\,
(y-y^{-1})^{-2} \nonumber \\ &&
\times \Big( y^{\alpha_{12}+\alpha_{13}+\alpha_{23}}
- y^{\alpha_{13}+\alpha_{23}-\alpha_{12}}
- y^{\alpha_{12}-\alpha_{23} -\alpha_{13}}
+y^{-\alpha_{12}-\alpha_{13}-\alpha_{23}} \Big)\, ,
\nonumber \\
\een
in agreement with the result \eqref{gref3}. Further checks for 
$n>3$ can be found in \cite{Manschot:2010qz}. It is an open problem
to show by combinatorial means 
that the result \eqref{eclassq} agrees with the KS formula 
in general.

\section{Wall-crossing from Abelian quivers \label{sec_qui}}

Finally, we discuss an alternative computation of the combinatorial factors 
$g_{\rm ref}(\{\alpha_i\}, y)$ based on Reineke's formula for invariants of
quivers without closed loop. 

The physical idea is that the supersymmetric quantum mechanics of
multi-centered black holes admits two branches: the Coulomb branch,
where the centers are far separated and well described by the supergravity
solution described in \S\ref{sec_sugra}, and the Higgs branch, where the
centers are very close to each other and are better represented as D-branes,
with open strings stretched between them. At large string
coupling, the wave function is mainly supported on the Coulomb branch,
while at small string coupling it is mainly supported on the Higgs branch  \cite{Denef:2002ru}. 
However, the BPS index $\Tr'(-1)^F$ is independent of the string coupling (and other
hypermultiplet fields), and should be computable in both. It is less clear that
the same should be true of the refined index $\Tr'(-y)^{2J_3}$, since this quantity 
is in general non-protected in string theory \cite{Gaiotto:2010be}, but for what 
concerns the jump of the refined index across walls of marginal stability, we shall
find strong evidence that this is the case. 

On the Higgs branch, the D-brane system is described at low energy by a 
supersymmetric quiver quantum mechanics, with gauge group $\prod_{i=1}^{n} U(N_i)$,
where $N_i$ is the number of coinciding D-branes at point $i$, and $\langle
\alpha_i,\alpha_j$ fields in the bifundamental representation $(N_i,\bar N_j)$
when  $\langle \alpha_i,\alpha_j\rangle$ is positive, or  $-\langle
\alpha_i,\alpha_j\rangle$ fields in the bifundamental representation $(\bar N_i,N_j)$
in the opposite case. Thanks  to the Bose-Fermi/Boltzmann correspondence
described in \S\ref{sec_stat}, we can treat all centers as distinguishable and 
assume that $N_i=1$ for any $i$, provided that we attach an effective rational 
index $\bOm_{\rm ref}(\alpha_i,y)$ to each center, and perturb the charge vectors
so that none of them coincide. Moreover, since the 
$\alpha_i$ can be ordered such that $\langle\alpha_i,\alpha_j\rangle\geq 0$
whenever $i<j$, the quiver admits no closed loop. 

For arbitrary quivers without loops, Reineke has computed the Poincar\'e polynomial
of the moduli space of quiver representations \cite{MR1974891}. In the special case
of Abelian quivers ($N_i=1$), Reineke's formula gives 
\be
\label{thebigformula}
 g_{\rm ref}(\{\alpha_i\}, y)
 =
 \frac{(-y)^{- \sum_{i<j} \alpha_{ij}} }{(y-1/y)^{n-1}}
 \sum_{\rm partitions}(-1)^{s-1} y^{2\sum_{a\leq b}\sum_{j<i }
 \alpha_{ji}\, 
 m^{(a)}_i \, m^{(b)}_j }\, ,
\ee
where the sum runs over all ordered partitions of 
$\gamma=\alpha_1+\cdots +\alpha_n$ into $s$ vectors
$\beta^{(a)}$ ($1\le a\le s$, $1\le s\le n$)
such that
\begin{enumerate}
\item$\beta^{(a)} = \sum_i m^{(a)}_i \alpha_i$ with
$m^{(a)}_i\in\{0,1\}$, $\sum_a \beta^{(a)} =\gamma$
\item 
$\langle \sum_{a=1}^b \, \beta^{(a)}, 
\gamma \rangle > 0 \quad
\forall \quad b \quad \hbox{with} \quad 1\le b\le s-1$
\end{enumerate}
We refer to \eqref{thebigformula} as the `Higgs branch wall-crossing formula'.

Let us illustrate how this formula works by
computing $g_{\rm ref}(\{\alpha_i\},y)$ for $n=2,3$.
For $n=2$ case with $\alpha_{12}>0$,
there are two possible ordered
partitions satisfying the conditions stated above:
\be \label{etwopart}
\{ \alpha_1+\alpha_2\}, \quad \{\alpha_1, \alpha_2\}
\, .
\ee
The first term contributes $y^{2\alpha_{12}}$ and the 
second
term contributes $-1$ to the sum. In total,
\be \label{erei2}
g_{\rm ref}(\alpha_1,\alpha_2,y) = (-y)^{1-\alpha_{12}}\, 
(y^2-1)^{-1} \, 
(y^{2\alpha_{12}} - 1)
= (-1)^{\alpha_{12}+1}\,
{y^{\alpha_{12}} - y^{-\alpha_{12}}
\over y - y^{-1}}\, ,
\ee
in agreement with \eqref{gref2}. For $n=3$,  
assuming the same conditions on $\alpha_i$ as in \eqref{gref3},
we find 6 possible ordered partitions
\be \label{erei4}
\{\alpha_1+\alpha_2+\alpha_3\}, \quad
\{\alpha_1, \alpha_2+\alpha_3\}, \quad
\{\alpha_1+ \alpha_2, \alpha_3\}, \quad
\{\alpha_1+ \alpha_3,\alpha_2\}, \quad
\{\alpha_1, \alpha_2, \alpha_3\}, \quad
\{\alpha_1, \alpha_3, \alpha_2\}\, .
\ee
The second and the
last contribution cancel, leaving
\be \label{gref3rei}
\begin{split}
g_{\rm ref}(\alpha_1,\alpha_2,\alpha_3, y)
=& (-1)^{\alpha_{12}+\alpha_{13}+\alpha_{23}}\,
(y - y^{-1})^{-2} \, \\ &
\left( y^{\alpha_{12}+\alpha_{13}+\alpha_{23}}
- y^{\alpha_{12}-\alpha_{23} -\alpha_{13}} 
- y^{\alpha_{13}
+\alpha_{23}-\alpha_{12}} 
+y^{-\alpha_{12}-\alpha_{13}-\alpha_{23})} 
\right)  \ ,
\end{split}
\ee
in agreement with \eqref{gref3}. 

\section{Conclusion and open problems\label{sec_conclu}}

In this survey, we have described four apparently different wall-crossing
formulae, the KS formula \eqref{ewallfinref}, the JS formula \eqref{egfromjs},
the `Coulomb branch' formula \eqref{eclassq} and the `Higgs branch' 
formula \eqref{thebigformula}. In \cite{Manschot:2010qz} we have checked that these
formulae agree among each other for $n\leq 5$. Unfortunately, we do
not have a mathematical proof of their equivalence at this point. 
The equivalence of the KS and JS formula appears  to be on solid 
ground \cite{ks,Joyce:2008pc}. The Reineke formula (or rather, its
specialization at $y=1$) is known to follow from the JS 
formula \cite{MR2357325} (although I do not know of an explicit 
combinatorial proof). The underlying mathematical structure of these
formulae involves Ringel-Hall algebras (and generalizations thereof \cite{ks2}), 
which might provide a realization of the long sought-after algebra of BPS 
states \cite{Harvey:1996gc,Fiol:2000wx}.

The Coulomb branch  formula \eqref{eclassq} appears 
to be new, and it would be desirable to derive it e.g. from the 
Higgs branch formula \eqref{thebigformula}. Indeed, the equivalence between these
two formulae could be viewed as a toy model of open/closed 
string duality. The comparison of the Higgs and Coulomb 
branch formulae suggests that the permutations $\sigma$ for which 
$s(\sigma)$ does not vanish are those whose maximal increasing 
subsequence\footnote{Rather than defining the notion of (maximal) 
increasing subsequence, we illustrate it on an example: for 
the permutation $\sigma(1234)=3142$, the increasing subsequences are
$\{\{3\},\{14\},\{2\}\}$ and $\{\{3\},\{1\},\{4\},\{2\}\}$, associated to the ordered
decompositions $\{\alpha_2,\alpha_1+\alpha_4,\alpha_3\}$ and 
$\{\alpha_2,\alpha_4,\alpha_1,\alpha_3\}$, respectively. The first increasing subsequence
is maximal, the second is not. More details can be found in \cite{Manschot:2010qz},   \S 3.3.}
satisfies the condition ii) below \eqref{thebigformula},
while none of its non-maximal increasing 
subsequence does, in which case $s(\sigma) = (-1)^{\#\{i; \sigma(i+1)<\sigma(i)\}}$.
It would be interesting to derive these conditions by applying Morse theory
to the `superpotential'  $W$ in \eqref{defhatW}. 

While the Coulomb and Higgs branch formulae have been derived in the context of
$\cN=2$ supergravity, the fact that they appear to agree with the KS and JS formulae 
suggests that they should work  just as well in the context of $\cN=2$ SYM theories.
For such field theories, unlike in supergravity, the refined index (or rather a variant
thereof,  known as  the protected spin character, and 
constructed out of the spatial rotation generator $J_3$ and a $SU(2)_R$ 
symmetry generator $I_3$ \cite{Gaiotto:2010be})
receives contributions from short multiplets only.
It would be desirable to derive the jump of the protected spin character
by quantizing multi-centered dyonic solutions of the low energy Abelian gauge theory, 
along the lines of \cite{Denef:1998sv,Lee:2011ph},
and see if it agrees with the Coulomb formula (presumably this jump will
coincide with the equivariant index of the Dirac operator $D$, without the 
need to assume that ${\rm Ker} D_-=0$). It would also be interesting to clarify
the relation with other realizations of BPS dyons in $\cN=2$ gauge theories based on 
string webs or non-Abelian monopoles \cite{Bergman:1997yw,Lee:1998nv,
Gauntlett:1999vc,Ritz:2000xa,Argyres:2001pv}.

More generally, localization techniques appear to be a powerful way
of quantizing multi-centered black hole solutions at fixed values of the moduli, 
not only in the vicinity
of walls of marginal stability. In contrast to the situation studied here, the 
phase space $\cM_n$ is in general no longer be compact, due to the 
presence of scaling solutions \cite{Denef:2002ru,Bena:2006kb,
Bena:2007qc,Denef:2007vg,deBoer:2009un}. We refer the reader 
to \cite{MPS2-in-progress} for an application of localization techniques
to this problem.

\ack I wish to thank J. Manschot and A. Sen for collaboration on the results reported in this 
note, and I. Bena, D. Joyce, D. Persson, S. El-Showk, M. Vergne for valuable discussions.  
I am also grateful to the organizers of the AGMP-06 meeting in Tj\"arn\"o for the invitation to present these results in a  stimulating and scenic environment.

\section*{References}

%\bibliography{../../HYPERTWI/common/combined}
%\bibliographystyle{iopart-num}
%
%\end{document}

\providecommand{\newblock}{}

\end{document}